\documentstyle[epsfig, twocolumn]{iso98}

\newcommand{\thepapername}{}

\begin{document}

\setlength{\parindent}{0pt}
\setlength{\parskip}{ 10pt plus 1pt minus 1pt}
\setlength{\hoffset}{-1.5truecm}
\setlength{\textwidth}{ 17.1truecm }
\setlength{\columnsep}{1truecm }
\setlength{\columnseprule}{0pt}
\setlength{\headheight}{12pt}
\setlength{\headsep}{20pt}
\pagestyle{veniceheadings}

\title{\bf THE ISO/NASA KEY PROJECT ON AGN SPECTRAL ENERGY
        DISTRIBUTIONS (CHARACTERISTICS OF THE ISO DATA) 
        \thanks{ISO is an ESA
project with instruments funded by ESA Member States (especially the PI
countries: France, Germany, the Netherlands and the United Kingdom) and
with the participation of ISAS and NASA.}}

\author{{\bf Eric Hooper$^1$, Belinda Wilkes$^1$, Kim McLeod$^2$, 
Martin Elvis$^1$, Chris Impey$^3$,} \\
{\bf Carol Lonsdale$^4$, Matt
Malkan$^5$, Jonathan McDowell$^1$} \vspace{2mm} \\
$^1$ Harvard-Smithsonian Center for Astrophysics, Cambridge, MA, USA \\
$^2$ Wellesley College, Wellesley, MA, USA \\
$^3$ Steward Observatory, Tucson, AZ, USA \\
$^4$ IPAC, Caltech, Pasadena, CA, USA \\
$^5$ UCLA, Los Angeles, CA, USA }

\maketitle

\begin{abstract}

The U.S. ISO Key Project on quasar spectral energy distributions seeks
to better understand the very broad-band emission features of quasars
from radio to X-rays.  A key element of this project is observations
of 72 quasars with the ISOPHOT instrument at 8 bands, from 5 to
200\,$\mu$m.  The sample was chosen to span a wide range of redshifts
and quasar types.  This paper presents an overview of the analysis and
reduction techniques, as well as general trends within the data set
(comparisons with IRAS fluxes, uncertainties as a function of
background sky brightness, and an analysis of vignetting corrections
in chopped observing mode).  A more detailed look at a few objects in
the sample is presented in Wilkes et al. (1999).
\vspace {5pt} \\


  Key~words: ISO; ISOPHOT; infrared astronomy; quasars; spectral
  energy distributions.

\end{abstract}

\section{INTRODUCTION}
\label{sec:intro}

A substantial fraction of the bolometric luminosity of many quasars
emerges in the infrared (Elvis et al. 1994), from synchrotron
radiation and dust.  Which of these emission mechanisms is dominant
depends on quasar type and is an open question in many cases.  Two
major ISO observing programs have obtained broad-band photometry for
large samples of quasars: a European Core program which focused on
low-redshift, predominantly radio-loud quasars; and a US Key Project
to examine quasars spanning a wide range of redshifts and SEDs, e.g.,
X-ray and IR-loud, plus those with unusual continuum shapes.

The final sample for the US Key Project consists of 72 quasars
observed with the ISOPHOT instrument (Lemke et al. 1996) in most or
all of the following bands: 5, 7, 12, 25, 60, 100, 135, and
200\,$\mu$m.  Ninety percent of the quasars in the sample have
redshifts $z < 1$, while the remaining 10\% lie in the range $2 < z <
4.7$ (see Hooper et al. 1999 for a plot of absolute blue magnitude
vs. redshift for the sample).  More than half of the sample consists
of luminous X-ray sources, 25\% are strong UV emitters, and smaller
subgroups contain strong infrared sources, X-ray-quiet objects, red
quasars, and BALQSOs.  The infrared data points will be combined with
all available fluxes at other wavebands to generate a comprehensive
atlas of broadband SEDs (see Wilkes et al. 1999 for some
examples).

Most of the sources (53 of 72) were observed in a rectangular chop
mode, the point source detection technique preferred at the beginning
of the ISO mission.  Concerns about calibrating and interpreting
chopped measurements, particularly at long wavelengths, led us to
switch to small raster scans.  We reobserved 18 of the chopped fields
in raster mode and added 19 new targets.  The change in observing
strategy, combined with lower than expected instrumental sensitivity,
has resulted in a halving of the originally planned sample.  However,
we now have the added benefits of data from both observing modes for a
subset of the targets and better information about background
variations from the raster maps.

In this paper we present an overview of the analysis and reduction
strategies and give an update of the status of the data products.
Reduction of faint object data taken with the ISOPHOT instrument has
been complex and somewhat uncertain, and the techniques are still in a
state of development.  Comparisons of our results with independent
checks, such as IRAS, and an overlap of the raster and chopped
observing methods help establish the validity of our data set, and the
large size of the sample provides a convenient testbed for a variety
of analysis procedures.

\begin{figure*}[!ht]
  \begin{center}
    \leavevmode
  \centerline{\epsfig{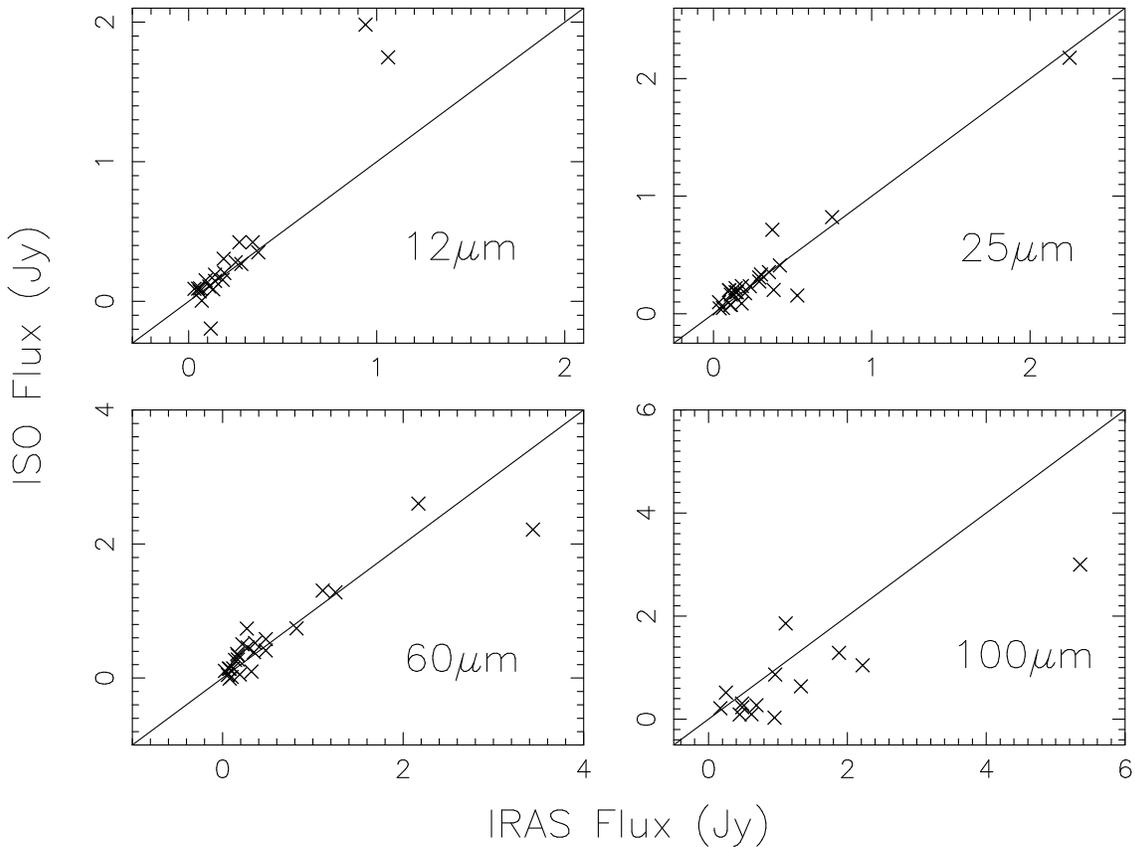}}
  \end{center}
  \caption{\em Comparison of ISO and IRAS fluxes for chopped
  measurements reduced with a Fourier transform code.
Lines corresponding to equal flux observed with both telescopes are provided
to guide the eye and do not represent any kind of fitting. }
  \label{fig:iso_iras}
\end{figure*}

\section{POINT SOURCE FLUXES}
\label{sec:basic_info}

The bulk of the data reduction, including most instrumental
calibrations and corrections, is done with PIA, the standard software
for ISOPHOT reductions (Gabriel, Acosta-Pulido, \& Heinrichsen 1998).
Basic steps are typically run in batch mode, including ramp
subdivision into 8 points per psuedo-ramp, two-threshold deglitching,
orbitally dependent dark current subtraction, and flux calibration
with the internal fine calibration sources, to produce AAP-level
files.  Standard instrumental drift corrections have not been employed
in chopped data, as there are too few points per chopper plateau.  We
have not yet explored the use of drift corrections in raster data.
The AAP files are left without vignetting corrections or sky
subtraction, as these steps are done with external custom scripts.

The final reduction steps, extracting the source flux and estimating
uncertainties, is an area of ongoing work, with multiple techniques
being explored.  To maintain flexibility, IDL scripts outside of PIA,
written by Martin Haas \& Sven M\"{u}ller and modified by us, are used
for both chopped and raster data.  We currently employ three main
techniques: a traditional source minus the average of adjacent
backgrounds for chopped data, from which are derived numerous
statistics; a Fourier analysis of the sequence of chopped
measurements, which is generally less affected by residual glitches
than the traditional approach but is more difficult to interpret; and
a simple average background subtraction in the raster maps to obtain
source flux and uncertainty estimates.  Many on-the-fly options are
available, including altering the vignetting values, finding and
correcting gaps in the chopper sequence, discarding part of the
sequence, flat fielding, background subtraction, plus plots of any
aspect of the data.

Fluxes derived using the Fourier transform analysis of chopped
measurements of relatively bright sources are compared to IRAS values
in Figure \ref{fig:iso_iras}.  The agreement is generally good;
possible explanations for the few large deviations include residual
glitches or flux calibration errors in ISO, errors in IRAS fluxes, or
intrinsic source variation.  A similar plot was presented in Hooper et
al. 1999 based on an earlier stage of the analysis techniques.  An
improvement in the agreement of the ISO and IRAS fluxes in the current
version is particularly apparent at 60\,$\mu$m.

\section{BACKGROUND ESTIMATES}
\label{sec:background}

Intrinsic sky structure noise can dominate instrumental uncertainties
in ISOPHOT measurements at wavelengths $\lambda \geq 100 \mu$m
(Herbstmeier et al. 1998).  Background fluctuations are particularly
problematic for simple chopped measurements with the four-pixel C200
array, where they contribute a systematic error that is difficult to
determine directly from the observations.  One possibility is to
estimate the structure noise using the results of Herbstmeier et
al. (1998), along with the prescription of Helou \& Beichman (1990) to
adjust for the observed sky brightness and selected wavelength.

\begin{figure}[!h]
  \begin{center}
    \leavevmode
  \centerline{\epsfig{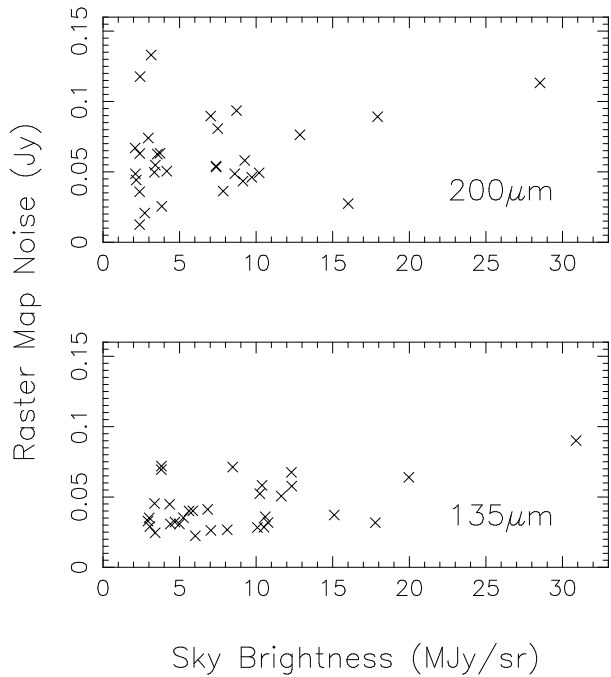}}
  \end{center}
  \caption{\em Rms noise vs. sky brightness for the C200 array raster
  data, calculated from off-source positions in the same manner as the
  flux for the source.  Includes contributions from both the
  instrumental noise and the sky structure noise.  }
  \label{fig:C200noise}
\end{figure}

We measure the background variations more directly where possible from
$2 \times 4$ raster scans with the C200 array, using the same
technique as the source flux determination.  The difference between
the flux at a sky position in the raster sequence and the average of
the preceding and following pointings (cases in which the source is
centered on the pixel in any of these three positions are excluded)
for each pixel forms a sequence of 12 measurements from which to
estimate the total noise contribution, including structure noise and
instrumental effects.  These noise figures are plotted against
measured sky brightness in Figure \ref{fig:C200noise} for most of the
raster maps in our sample.  A weak trend of increasing noise with sky
brightness is evident at 200\,$\mu$m.  The results are generally
consistent with Herbstmeier et al. (1998), given that extrapolations
of the structure noise between different fields and brightness levels
can differ from the measured values by a factor of 5 or more.

\section{VIGNETTING CORRECTIONS}
\label{sec:vignetting}

Vignetting corrections are important parameters for chopped faint
source observations, as relatively small errors can produce large
changes in the computed source flux.  Working with Martin Haas, we
derived vignetting corrections from our C200 chopped data (135 \&
200\,$\mu$m) to compare with the default values.

The first step was to calculate average flat fields for the array in
the on and off-source positions separately.  Given the large sample
size, we expect that any observed differences in the measured flat
fields reflect changes in the vignetting between the two chopper
positions.  With proper vignetting corrections applied, the values
should be close for each pixel.  This was not the case for the default
corrections; the flat fields differed by a similar or larger factor
than with no correction applied at all.

The ratios of the on and off-source flat fields were used to estimate
vignetting corrections, a process aided by the geometry of the array
and the chopper motion.  All of the observations had a rectangular
chopping pattern of $\pm 90$ arcsec, a total throw approximately equal
to the projected angular size of the array.  The spacecraft pointed
halfway between the source and the background field.  In this
configuration, one half of the array imaged the central field of view
of the spacecraft in each chopper position.  Assuming that the central
field had relatively uniform low-level vignetting (in the calculations
it is set to 1.0, the same value used for the central field with no
chopper movement), the flat field ratios directly gave the vignetting
correction factors for the outer part of the chopping pattern.  The
derived corrections, listed in Table \ref{tab:vignet}, are smaller and
more uniform than the standard values, which range from 1.01 to 1.10.
In addition, the new numbers are smaller at the longer wavelength,
whereas many but not all of the default corrections follow the
opposite trend.  We are still evaluating whether the new or standard
vignetting corrections are closer to the true values.

\begin{table}[h]
  \caption{\em  Vignetting corrections for the C200 array with a
  total chopper throw of 180 arcsec.  Derived from the average flat fields
  of each pixel in the on-source (+90 arcsec chopper position) and
  off-source (-90 arcsec chopper position) configurations.    }
  \label{tab:vignet}
  \begin{center}
    \leavevmode
    \footnotesize
    \begin{tabular}[h]{lcccc}
      \hline \\[-5pt]
      $\lambda$ \& chop position & Pixel 1 & Pixel 2  &  Pixel 3  &  Pixel 4 \\[+5pt]
      \hline \\[-5pt]
      135\,$\mu$m on source          & 1.018  & 1.000  & 1.000  & 1.024  \\
      135\,$\mu$m off source         & 1.000  & 1.026  & 1.015  & 1.000  \\
                                     &        &        &        &        \\
      200\,$\mu$m on source          & 1.009  & 1.000  & 1.000  & 1.010  \\
      200\,$\mu$m off source         & 1.000  & 1.022  & 0.999  & 1.000  \\
      \hline \\
      \end{tabular}
  \end{center}
\end{table}

\section*{ACKNOWLEDGMENTS}

Martin Haas, Sven M\"{u}ller, Mari Poletta, and Ann
Wehrle provided invaluable help with the data reduction.  We
benefitted greatly from discussions with Thierry Courvoisier,
P\'{e}ter \'{A}brah\'{a}m, Ilse van Bemmel, Rolf Chini, and Bill
Reach.  The IDC, IPAC, and the INTEGRAL Science Data Center were very
hospitable during visits to work on this project.  The financial
support of NASA grant NAGW-3134 is gratefully acknowledged.

\end{document}